\documentclass[10pt, twocolumn]{IEEEtran}
\usepackage{cite,graphicx,amsmath,amssymb}
\usepackage{subfigure}
\usepackage{citesort}
\usepackage{fancyhdr}
\usepackage[table]{xcolor}
\usepackage{array}
\usepackage{mdwmath}
\usepackage{mdwtab}
\usepackage{balance}
\usepackage{xcolor}
\usepackage{bm}
\usepackage{amsthm}
\usepackage{amssymb}
\usepackage{threeparttable}
\usepackage{algorithm}
\usepackage{algorithmic}
\usepackage{multirow}
\usepackage{flafter}

\begin{document}
\title{Integrated 3C in NOMA-enabled Remote-E-Health Systems}
 %\markboth{\textit{A Manuscript Submitted to The IEEE Wireless Communications} }{}
\author{
Xiao~Liu,~\IEEEmembership{Student Member,~IEEE,}
 Yuanwei~Liu,~\IEEEmembership{Senior Member,~IEEE,}\\
   Zhong~Yang,~\IEEEmembership{Student Member,~IEEE,}
     Xinwei~Yue,~\IEEEmembership{Member,~IEEE,}\\
     Chuan~Wang,
        and Yue~Chen,~\IEEEmembership{Senior Member,~IEEE,}

%\thanks{

%X. Liu, Y. Liu, Z. Yang and Y. Chen are with the School of Electronic Engineering and Computer Science, Queen Mary University of London, London E1 4NS, UK. (email: x.liu@qmul.ac.uk; yuanwei.liu@qmul.ac.uk; zhong.yang@qmul.ac.uk; yue.chen@qmul.ac.uk).}

%\thanks{X. Yue is with the School of Information and Communication Engineering, Beijing Information Science and Technology University, Beijing 100101, China. (email: xinwei.yue@bistu.edu.cn).}

%\thanks{C. Wang is with Capital Medical University Affiliated Beijing Anzhen Hospital, Department of Cardiovascular Surgery, Beijing 100029, China. (email: ts\_kevin@hotmail.com).}

}
 \maketitle

\begin{abstract}

A novel framework is proposed to integrate communication, control and computing (3C) into the fifth-generation and beyond (5GB) wireless networks for satisfying the ultra-reliable low-latency connectivity requirements of remote-e-Health systems. Non-orthogonal multiple access (NOMA) enabled 5GB network architecture is envisioned, while the benefits of bringing to the remote-e-Health systems are demonstrated. Firstly, the application of NOMA into e-Health systems is presented. To elaborate further, a unified NOMA framework for e-Health is proposed. As a further advance, NOMA-enabled autonomous robotics (NOMA-ARs) systems and NOMA-enabled edge intelligence (NOMA-EI) towards remote-e-Health are discussed, respectively. Furthermore, a pair of case studies are provided to show the great performance enhancement with the use of NOMA technique in typical application scenarios of 5GB in remote-e-Health systems. Finally, potential research challenges and opportunities are discussed.

\end{abstract}

\section{Introduction}

With the rapid economic development and improvement of living standards, the high-quality demands of medical health and care are gradually increasing. The full set of qualified health services are provided in highly urbanized areas, which are very well covered by communication networks. However, it is non-trivial to cover the sparsely populated rural areas. The fifth-generation and beyond (5GB) wireless networks have evoked a great deal of attention, due to its ability to reflect a large diversity of communication requirements and application domains\cite{Saad20196G,Liu8114722Beyond}. The 5GB is more ambitious than only supporting ultra-reliable and secure communication links, but also aims for providing uninterrupted and ubiquitous connectivity, which meet the requirements of remote-e-Health systems such as 5G platform based tele-medicine/tele-surgery, autonomous robotics enabled medical resource delivery and mobile health monitoring\cite{Di20195Ghealth}.

\begin{figure*} [t!]
\centering
\includegraphics[width=6in]{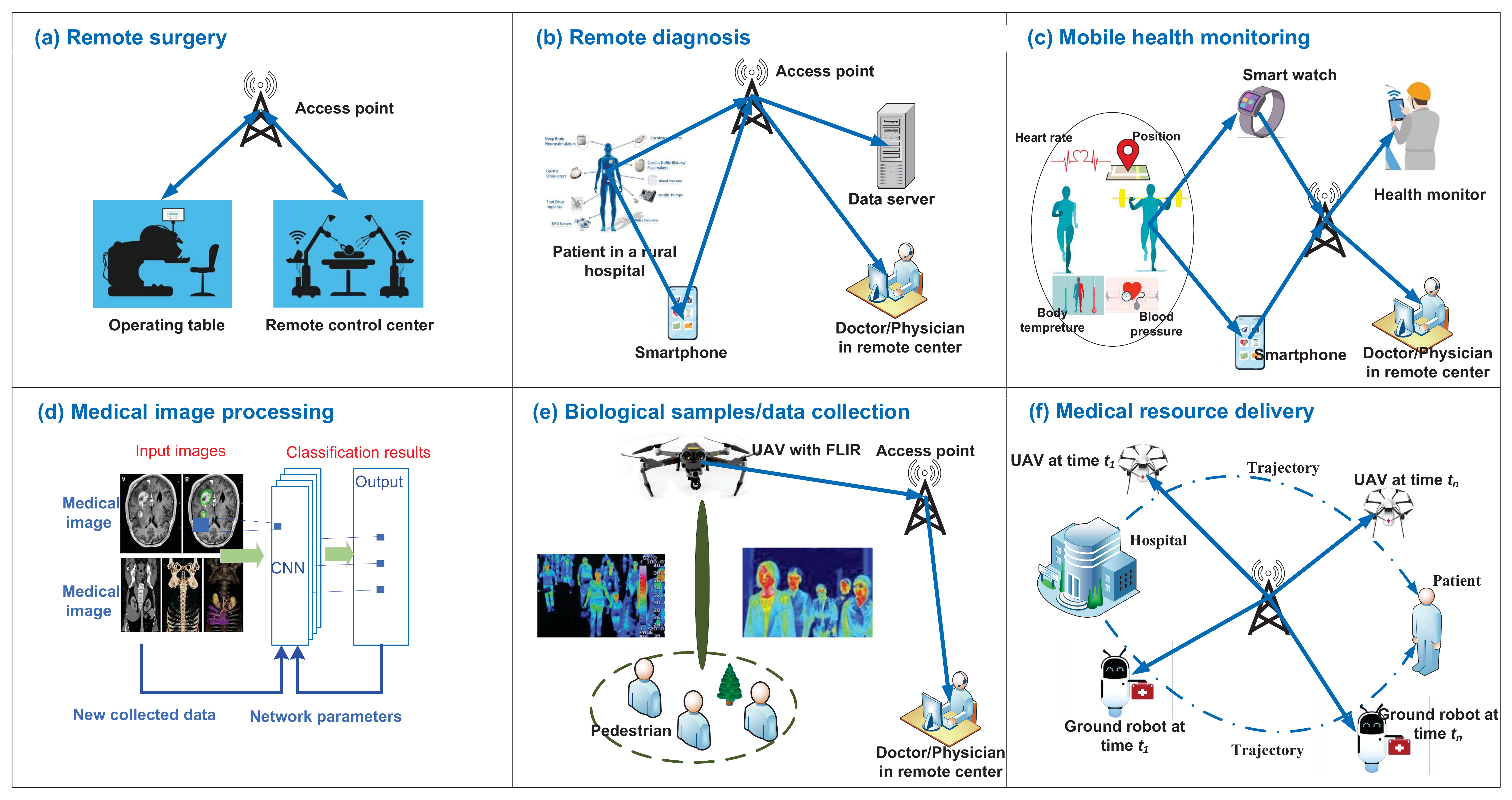}
\caption{Applications of 5GB in remote-e-Health systems.}\label{3C}
\end{figure*}

Fig. \ref{3C} illustrates several typical applications of 5GB in remote-e-health systems. In Fig. \ref{3C}(a), a remote surgery can be operated safely and smoothly with the aid of 5GB cellular networks. Thus, physicians in hospitals of highly urbanized areas can perform an operation on patients in areas where Internet cables are difficult to lay or cannot be laid. Fig. \ref{3C}(b) characterizes the application of 5GB in remote diagnosis. Real-time data derived from electroencephalography and electrocardiography can be transmitted from a rural hospital or community clinic to a remote center, where online diagnosis are made by experienced physicians. Fig. \ref{3C}(c) shows how physicians in a remote center can monitor individuals' health condition by invoking smart-phones and smart-watches to record their position and physical data (e.g., heart rate, blood pressure, body temperature) and transmit these data to physicians via 5GB cellular networks. In Fig. \ref{3C}(d), medical images, which are usually with jumbo size and are often regarded as burdens for cloud networks, can be processed by offloading computing resources to edge devices via 5GB networks, since 5GB networks are expected to carry a significantly higher volume of data while maintaining reliability. Fig. \ref{3C}(e) illustrates the employment of autonomous robotics (ARs) to collect individuals' biological data. The 5GB networks enables medical groups or related organizations to collect pedestrians's body temperature through forward looking infrared radar (FLIR) empowered unmanned aerial vehicles (UAVs), or collect samples of individuals who are self-isolated. In Fig. \ref{3C}(f), cellular-connected ARs are employed for delivering medical resources from hospital to individuals in a time-saving and energy-saving manner.

Non-orthogonal multiple access (NOMA), whose key idea is to superimpose the signals of two users at different powers for exploiting the spectrum efficiently by opportunistically exploring the users' different channel conditions, is capable of satisfying the requirements of massive connectivity, as well as high access speed and low latency in remote-e-Health systems. The goal of this article is to provide a potential solution to realize the application of 5GB in remote-e-Health systems through the NOMA technique. Before fully reap the benefits of NOMA-enabled 5GB networks in remote-e-Health systems, several research challenges have to be tackled, ranging from investigating the performance limits of NOMA-enabled 5GB networks, jointly designing the control policy and resource allocation policy for ARs, as well as determining the computing resource offloading policy in remote-e-Health systems. Motivated by the aforementioned challenges, the main contributions of this articles are integrating communication, control and computing (3C) in 5GB-enabled remote-e-Health systems. In contrast to the recent research contributions, we highlight how NOMA technique copes with the challenges and improves the performance of remote-e-Health systems. The application of the proposed NOMA-enabled 5GB networking framework is discussed in the context of NOMA-AR and NOMA-mobile edge computing (MEC) case studies.

\section{NOMA enabled remote-e-Health systems}
In this section, we first describe the NOMA-enabled 5GB remote-e-Health structure. Then the application of a unified NOMA framework is introduced to the remote-e-Health systems.
\subsection{NOMA for remote-e-Health}
\begin{figure*}[t!]
\centering
\subfigure[NOMA-enabled remote-e-Health structure.]{
\label{Mg_system_model}
\includegraphics[width=3.1 in]{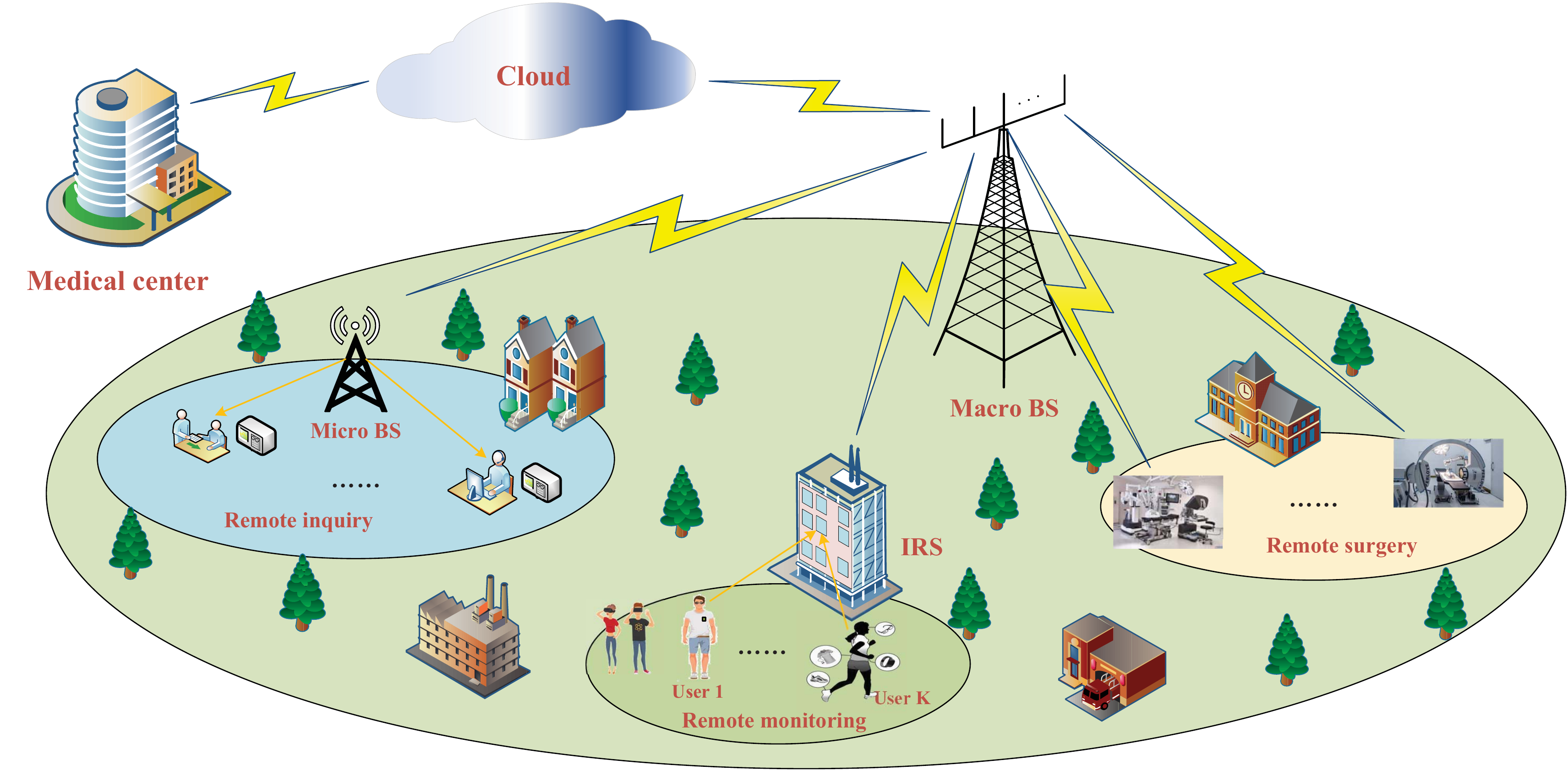}}
\subfigure[A unified NOMA for remote-e-Health systems.]{
\label{Uplink NOMA for remote-e-Health system}
\includegraphics[width=2.5 in]{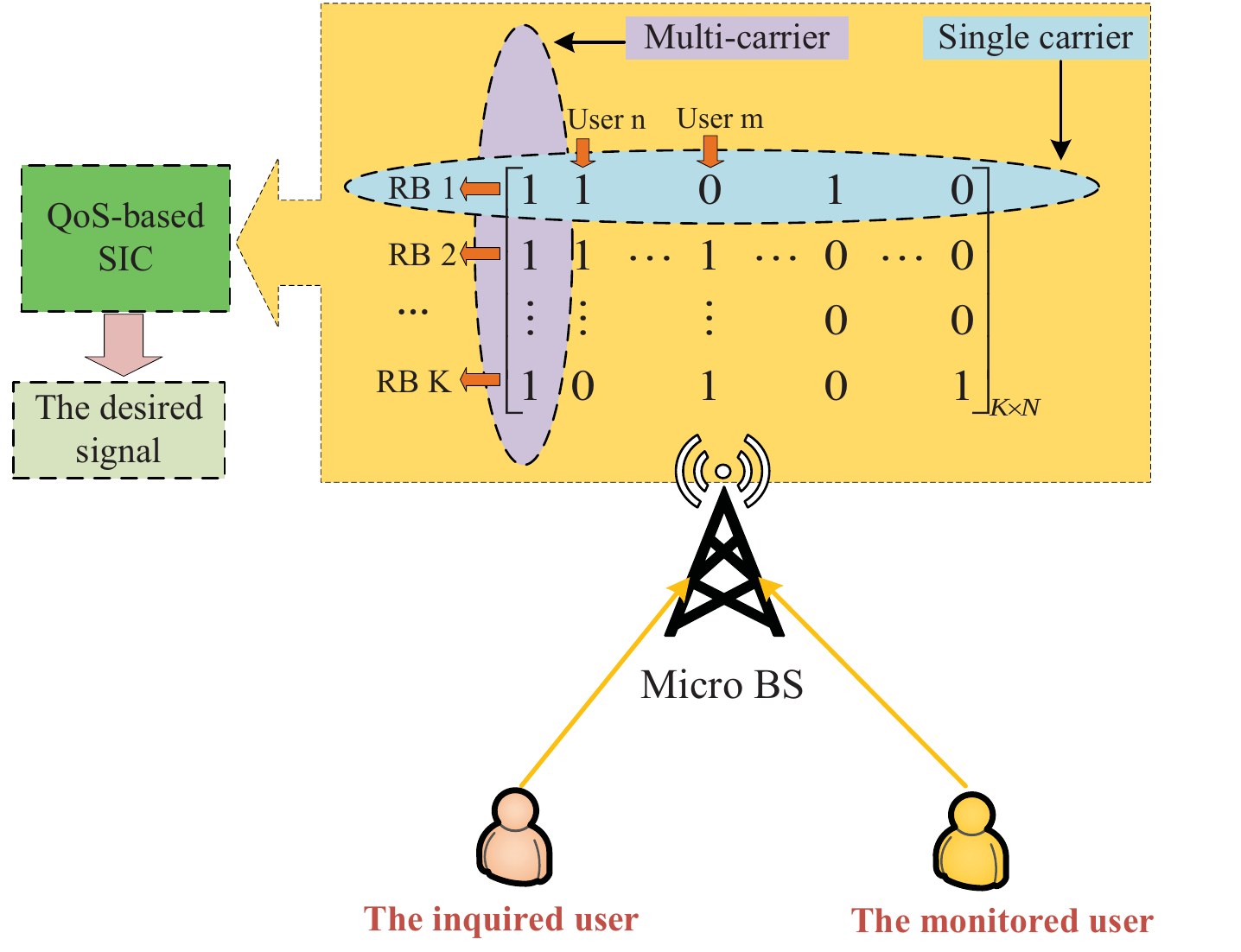}}
\caption{NOMA enabled remote-e-Health systems.}
\label{Fig. 1}
\end{figure*}

Due to supporting the enhanced spectrum efficiency and massive connectivity, NOMA technology has received an increased attention for the rising emergence of remote e-Healthcare services in the recent years. The remote-e-Health is an effective approach to solve the lack of health infrastructures and resources for the community, suburban and remote territories. The focus on the interoperability of personal health devices has intensified especially in the context of medical healthcare environments, which is capable of supporting the habitants of remote locations, chronic patients and independent living of elderly people. Currently, the remote-e-Health is still in the development stage, and it needs to satisfy the requirements of larger connectivity and individualization. Hence the application of NOMA to remote-e-Health system is one of better choices for the development of healthcare. As shown in Fig. \ref{Mg_system_model}, the total remote-e-Health network is deployment of multiple cells (e.g. macro, micro and intelligent reflecting surface (IRS) etc.). In NOMA-enabled remote-e-Health system, the NOMA protocol is adopted in each small cells and massive multiple-input and multiple-output technologies are employed by macro cells. For the macro cells, base stations (BSs) are equipped with multiple antennas and transmit the signals to carry out the remote surgery for patients. Meanwhile, the macros can be connected to medical center by utilizing optical fiber or wireless cloud networks. In particular, assuming that the transmitting antennas of macro BSs is much larger than the number of users. For the micro cells, users are equipped with single-antenna monitoring device and are allocated with the different power levels to carry out NOMA protocol. In the following subsection, we will discuss a unified NOMA frameworks for remote-e-Health in detail.

\subsection{A unified NOMA framework for remote-e-Health}
The superior spectral efficiency of NOMA has been demonstrated by extensive theoretic studies as well as experimental trials executed by academia and industry. Up to now, multiple NOMA proposals have been examined by different industrial companies \cite{Ding2017Mag}. Based on spreading signature of proposals, NOMA schemes can be classified as two categories: power-domain NOMA (PD-NOMA) and code-power NOMA (CD-NOMA). The superposed signals of multiple users are mapped into a single resource element (RB) for PD-NOMA, while the superposed signals are mapped into multiple subcarriers for CD-NOMA. Based on this property, a unified NOMA framework was discussed in \cite{Qin8387207UDN}, which can be reduced into PD/CD-NOMA according to the changes of scenario and business. The remote-e-Health combined with the unified NOMA framework is further capable of supporting its diversified remote businesses flexibly. As shown in Fig. \ref{Uplink NOMA for remote-e-Health system}, we take the uplink PD-NOMA enabled remote-e-Health as an example. Assuming that a pair of  remote users are served by the PD-NOMA scheme, where the medical center needs to monitor or inquiry elderly people's health profile in their own home. According to the requirements of quality of service (QoS), the inquired user i.e., the $n$-th user is seemed as a delay-sensitive user with a low target data rate, which needs to send the urgent health status changes. On the other hand, the monitored user i.e., the $m$-th user is served in a delay-tolerant mode with sending the personal health records. Finally, the QoS-based successive interference cancellation (SIC) scheme is carried out to first detect the inquired user's signal and then decode the monitored user's information. Certainly this procedure can also be easily extend to general case with more than two remote users, while ensuring the users' QoS requirements.

\section{NOMA-enabled Autonomous Robotics for Remote-E-Health: Integrated Communication and Control}

To validate the distinguished capabilities of NOMA for remote-e-Health systems, a NOMA-enabled autonomous robotics architecture is presented in this section.

\subsection{Key features of NOMA-enabled autonomous robotics for remote-e-Health}

Autonomous robotics, which has brought tremendous changes in various socio-economic aspects in our society, has also been the focal point of the e-Health research field. Autonomous robots are partitioned into three categories: aerospace robot (e.g., small intelligent satellite, airship/airplane, high-altitude platform (HAP), and unmanned aerial vehicle); ground robot (e.g., autonomous vehicle (AV), smart home robot, and mobile robot); marine robot (e.g., unmanned ship, unmanned submarine, underwater robot)\cite{Tzafestas2020robotics}.

\begin{figure*} [t!]
\centering
\includegraphics[width=4in]{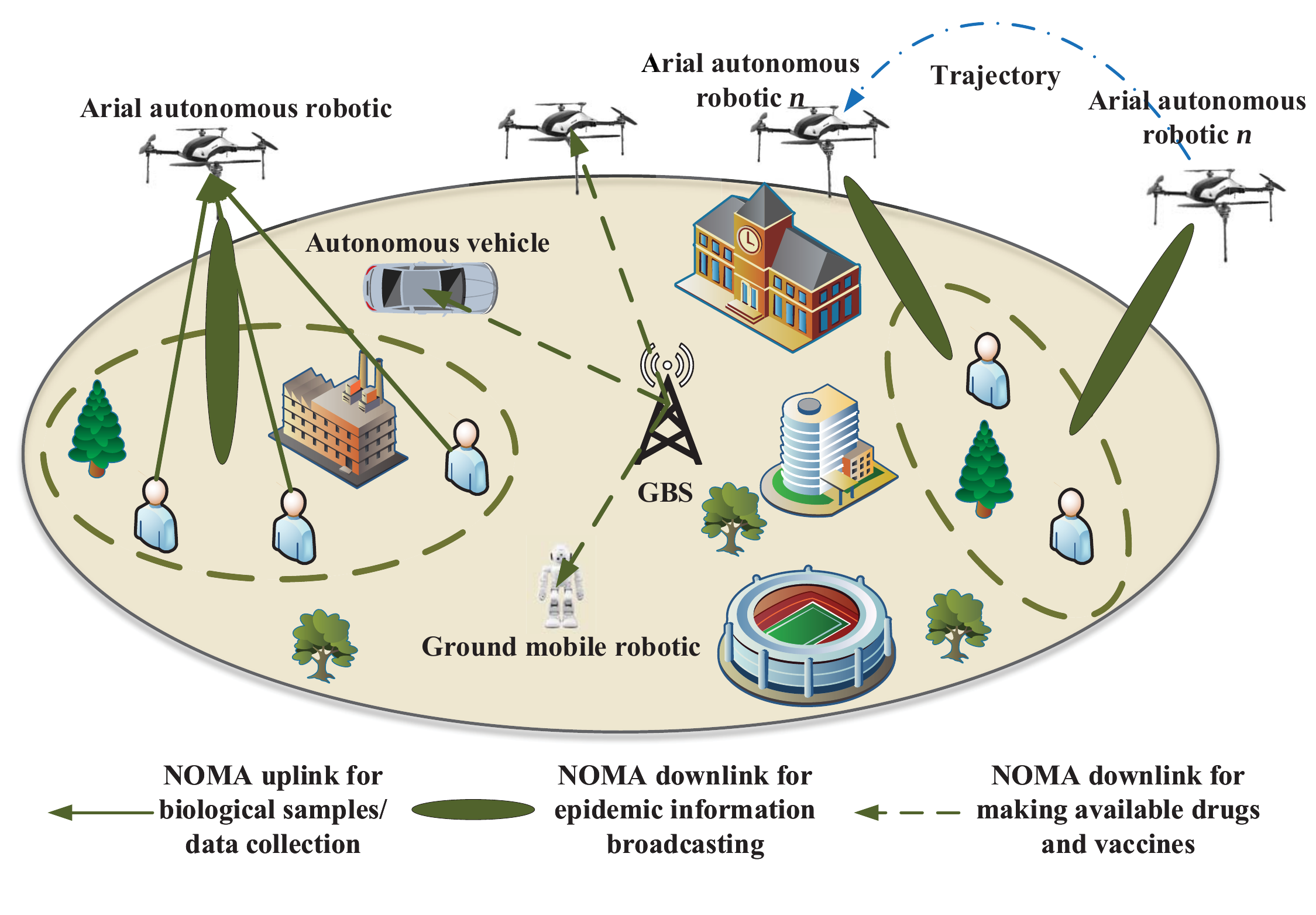}
\caption{NOMA-enabled autonomous robotics for 5GB remote-e-Health.}\label{ARscenario}
\end{figure*}

In previous research contributions, ARs are found to be extremely useful to replace humans or manned aircrafts/vehicles in missions that are dull (e.g., extended surveillance), dirty (e.g., pesticide spray), and dangerous (e.g., rescue and search after disaster)\cite{liu2019trajectory}. With the aid of ARs, revolutionary benefits have been witnessed in the area of e-Health. As shown in Fig.~\ref{ARscenario}, UAVs and ground mobile robots are already used for medical service delivery and biological information monitoring, while ground social robots are used for dealing with the mental health. ARs can make decision without the intervene from human, because that they can receive reliable real-time traffic information from the cellular network via access point (AP)/BS-ARs wireless links. It is worth mention that ARs invoked in e-health are danger-intolerable and delay-sensitive. It means that it is of the upmost importance to guarantee that ARs can move safely, which requires ultra-reliable and secure communication links for them. To provide uninterrupted and ubiquitous connectivity for ARs, as well as provide massive connectivity in ARs networks, power-domain NOMA can be adopted to support different users over the same time/frequency slot by exploring their different channel conditions.

\subsubsection{NOMA-enabled autonomous robotics for making available drugs and vaccines}

UAVs/AVs-based delivery systems have attracted remarkable attention in recent years due to the reason that a faster delivery speed and a lower cost can be obtained than the conventional manned delivery services, especially when the destination areas are inaccessible. In remote-e-Health system, ARs can be invoked for making available drugs, vaccines and other medical resource to un-reached individuals. During the delivery process, critical control information has to be transmitted from BSs/APs to ARs, which requires the communication signal to be ultra-reliable and low-latency. In the NOMA-enabled ARs systems, ARs with different channel conditions can be partitioned into a same NOMA cluster. Additionally, a particular AR can also be partitioned into the same cluster with other cellular network users. Since the wireless service quality for ARs has to be guaranteed at each timeslot, the trajectory of ARs needs to be designed based on the radio map. The movement of ARs affects the decoding order among ARs and user clustering, which makes the decoding
order design, user clustering and resource allocation highly coupled in the NOMA-enabled ARs systems.

\subsubsection{NOMA-enabled autonomous robotics for collecting biological samples/data}

In the previous research contributions, UAVs/robots are capable of collecting data from ground Internet of Things (IoT) devices in a given geographical area, in which constructing a complete cellular infrastructure is unaffordable. Sparked by this application, ARs are capable of being used for collecting biological samples/data from small sensors carried by individuals, which are used to monitoring individuals physical conditions (e.g., heart rate, blood pressure, body temperature) in both outdoor and indoor scenarios. Since these sensors/devices cannot support long distance transmission due to their power constraints, ARs can be helpful to collect the data and then efficiently transmit the data to other linked devices or directly to the data center. Additionally, ARs can also be invoked to collect individuals' biological samples in an efficient and non-touch manner. Given the position of all individuals, ARs are capable of automatically designing the most energy-efficient trajectory or fulfilling tasks within minimal flight duration. Since aerial robotics are battery-powered, their energy consumption is one of the gravest challenges. The limited flight-time of aerial robotics (usually under 30 minutes) hampers the wide commercial roll-out of AR-aided networking. Hence, improving the transmission efficiency from sensors to ARs is of utmost importance. With the aid of NOMA technique, the reliability of the wireless link can be improved due to superior spectrum efficiency of NOMA-aided networks. The information age can be minimized by jointly optimizing the sub-channel assignment, the decoding order, the uplink transmit power of medical nodes, as well as ARs' trajectory.

\subsection{NOMA for autonomous robotic communication: a machine learning approach}

We consider the multiple aerial autonomous robotics aided wireless networking scenario, where multi-ARs are invoked as aerial base stations for epidemic information broadcasting with the aid of NOMA technique. It is assumed that all ARs are deployed in the cellular utilizing the same frequency band, as a result, inter-cell interference has to be considered in this multi-cell network. Ground users are individuals who would like to receive real-time epidemic information in both video and text formats. Ground users are considered as roaming continuously instead of keeping static. Therefore, ARs have to be re-deployed based on the movement of users for guaranteeing high quality of wireless services. Intending to maximize the total throughput, we optimize the ARs's trajectory and power allocation
policy, subject to the maximum power constraint, decoding order constraints, and the QoS constraint. It is worth noting that the dynamic decoding order has to be determined at each time slot to guarantee the successful SIC, since the position of both UAVs and users is time-varying.

Users are typically considered to be stationary in the conventional convex optimization based ARs-aided wireless networks and the data demand of users is also considered to be time-invariant, hence only static communication environments are considered in conventional ARs-aided wireless networks. Hence, the control policy can indeed be analyzed based on tractable models, but ARs are not capable of learning from the environment or from the feedback of the users for enhancing their ability to cope with the dynamically fluctuating propagation and tele-traffic environment. In an effort to tackle the aforementioned challenges, reinforcement learning (RL) algorithms are adopted due to the reason that they are capable to indicate a near-optimal solution through an experience-based method but not a functional express. The ARs accumulate certain experiences through continuous exploration of the current environment and obtains high reward solutions by learning and remembering these fruitful or dreadful experiences.

The advantage of online RL model is that it can adapt the control policy of autonomous robots to the dynamic environment in real-time. However, online RL model requires additional communication resource for uploading each autonomous robot's real-time mobility information or feedback to base stations. In contrast to online RL model, offline RL model does not require additional communication resource, user mobility/QoS requirement can be predicted offline. However, performance loss cannot be avoid compared to online RL model. Since ARs are battery-powered, their energy consumption is one of the gravest challenges, especially when on-board data processing and on-line computation are used.

\begin{figure} [t!]
\centering
\includegraphics[width=3in]{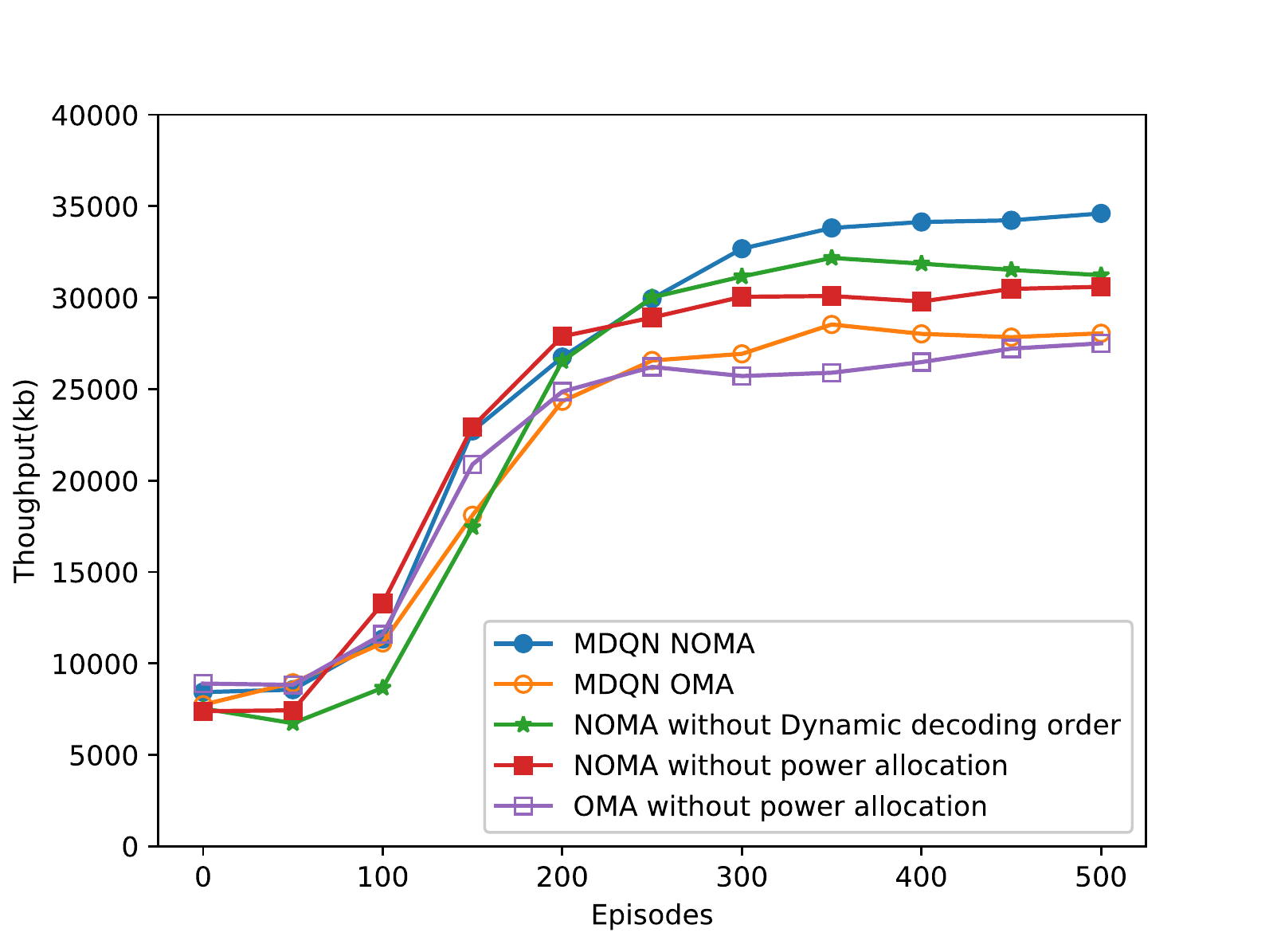}
\caption{Throughput enhancement of NOMA in ARs-aided wireless networks.}\label{ARs}
\end{figure}

Fig.~\ref{ARs} characterizes the throughput of the network over episodes. The results are derived from both the deep Q-network (DQN) algorithm and the mutual deep Q-network (MDQN) algorithm, in which multiple ARs are permitted to connect with the same neural network during the training process with the assistance of state abstraction. One can observe that the NOMA-enabled ARs network is capable of achieving roughly 23\% enhancement than OMA case in terms of throughput. It can also be observed that in the NOMA case, the dynamic decoding order and the power allocation policy derived from the proposed MDQN algorithm are capable of achieving gains of approximately 12\% and 14\%.

\section{NOMA-enabled Edge Intelligence for 5GB Remote-E-Health: Integrated Communication and Computing}

In this section, we present the network structure of the NOMA-enabled edge intelligence (NOMA-EI) framework in remote-e-Health under consideration. Subsequently, the associated key modules are detailed.

\subsection{Network structure for NOMA-EI in remote-e-Health}

Recent proliferation of edge smart devices in remote-e-Health have been growing exponentially, e.g., smart watches or other portable sensing equipments. The prosperity of health services in remote-e-Health require the integration of communication and computing resource in network edge. To fulfill these requirements, EI is proposed to push network functions from the network centre to the network edge~\cite{Zhou2019ei}.

\begin{figure*}
\setlength{\abovecaptionskip}{-0.2cm}
  \centering
  \includegraphics[width=4.5in]{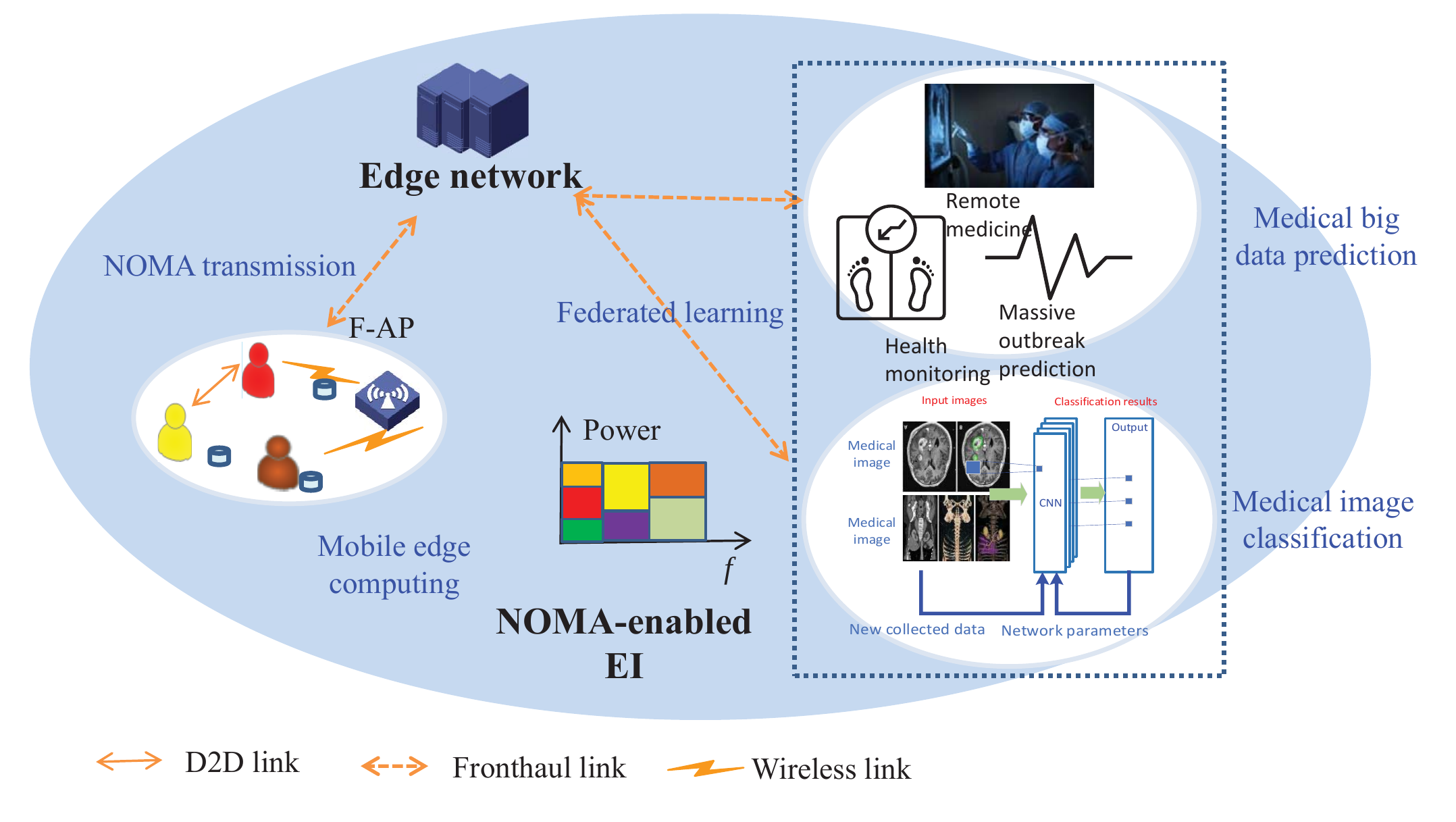}\\
  \caption{NOMA-enabled Edge Intelligence Framework in Remote-E-Health. (D2D denotes device-to-device, F-AP represents fog access point, SAQ-learning is single-agent Q-learning)}\label{framework}
\end{figure*}

Fig.~\ref{framework} illustrates the network structure of NOMA-enabled EI. NOMA strategy has shown great potential for circumventing the limitation of massive connectivity compared to conventional OMA techniques~\cite{Liu8114722Beyond}, it is anticipated that the unprecedented high access speed and low latency requirements could be well addressed by NOMA enabled EI. The marriage of NOMA and EI has given rise to a new research area, namely, NOMA-EI. In NOMA-EI, the data offloading between edge devices and edge server can be conducted by NOMA strategy, to fulfill the stringent data rate and delay requirements of data offloading and task computation using the limited energy and computation resource.

In NOMA-EI, high-computation capabilities is required for edge devices to fulfill the computing demand for remote-e-Health. To meet this demand, MEC is proposed to push the computation function of the network centre to the network edge of remote-e-Health. The motivation of NOMA-MEC for remote-e-Health is to take advantages of the massive connectivity characteristic of NOMA and the delay reducing property of MEC. The computation services requested by smart devices are transmitted to the BSs for emancipating smart equipments from onerous computing and reducing computation latency. The core of MEC is to promote plentiful computing capabilities at the edge of networks by integrating MEC servers at BSs. The NOMA-enabled MEC (NOMA-MEC) is capable of offloading the tasks simultaneously to reduce task transmission delay. This motivated extensive researches from both academia and industry. In~\cite{Liu2019ieeewc}, a novel framework of cooperative NOMA-MEC is proposed to exploit parallel task transmission on the whole NOMA transmission resource block.

\subsection{Reinforcement learning for NOMA-MEC in remote-e-Health}

The motivation of using RL for NOMA-MEC in remote-e-Health is to obtain a long-term offline solution for the optimization problem. The advantage of RL is that the agent of RL, differs from supervised learning algorithms, such as deep neural networks, do not needing labelled input/output pairs during the training process, therefore, the patients' privacies are preserved. The application of RL has been widely employed in various scenarios to obtain a best trajectory or path for a given environment. The aim of RL in remote-e-Health systems is to obtain the best trajectory from the experience of intelligent agent, and measure the best state-action value function in the environments of remote-e-Health. The advantage of RL is that the intelligent agent does not estimate all the state-action pairs, which is capable of reducing the computation complexity. By maximizing the long-term reward of the intelligent agent in NOMA-MEC, the associated stochastic optimization problems in NOMA-MEC are solved. However, the performance of the conventional RL is restricted by the dimensions of the action space and state space.

\begin{figure*}[t!]
\centering
\subfigure[The structure of DRL for NOMA-MEC in remote-e-Health.]{
\label{MEC}
\includegraphics[width=2.8 in]{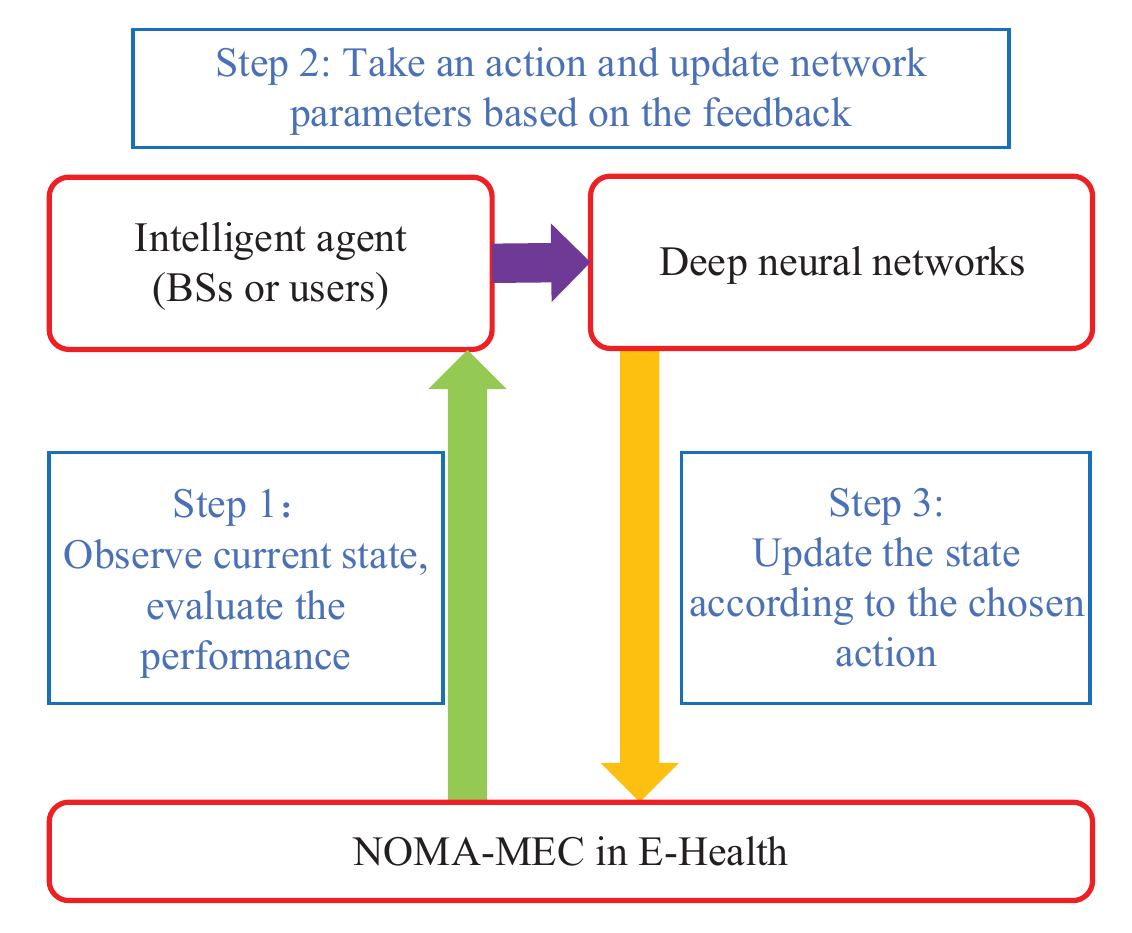}}
\subfigure[The performance of the proposed NOMA-MEC framework in remote-e-Health on the energy consumption~\cite{zhong2020twc}.]{
\label{DRLframework}
\includegraphics[width=3.1 in]{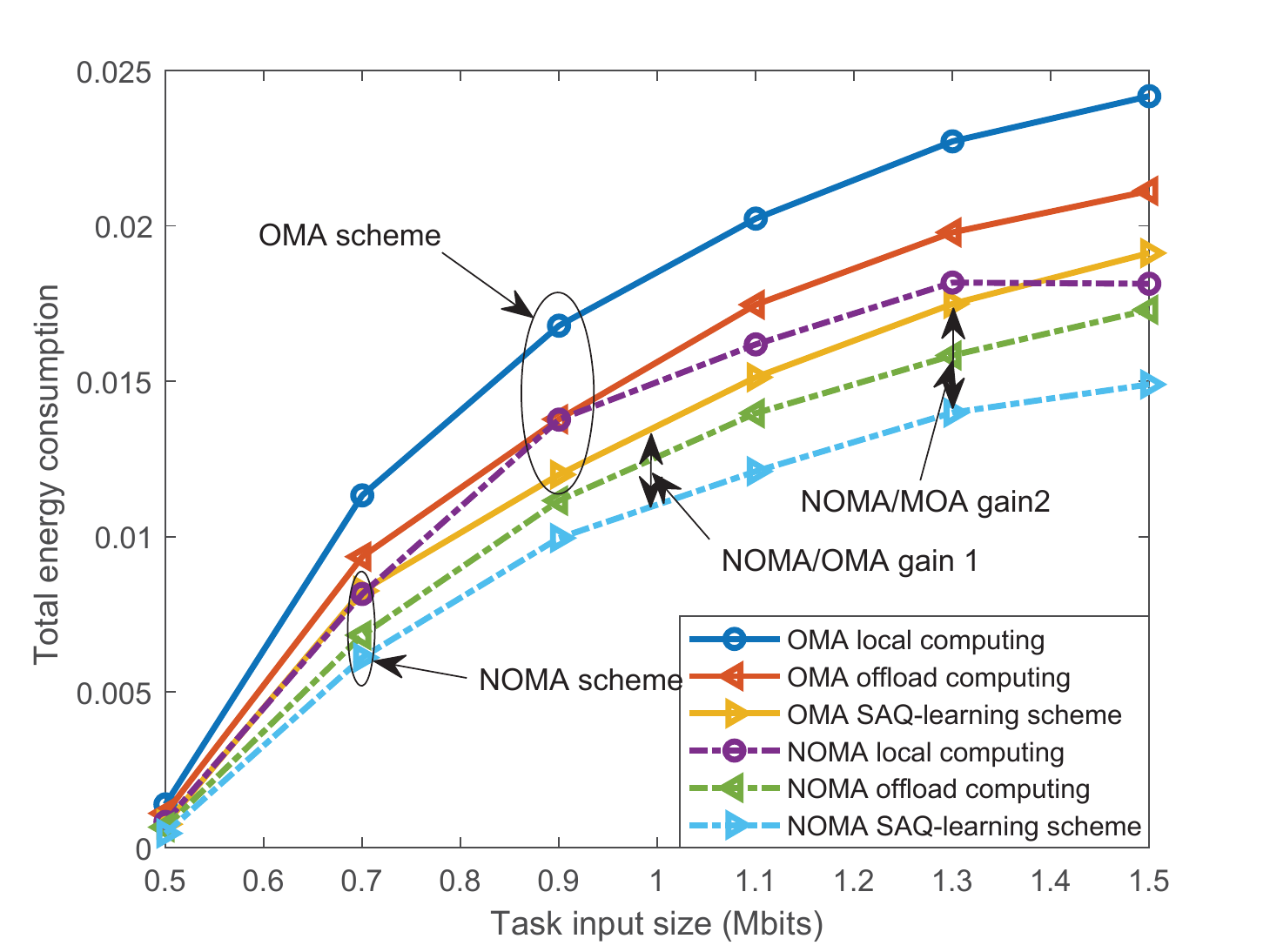}}
\caption{NOMA-enabled MEC for remote-e-Health.}
\label{offloading}
\end{figure*}

In order to overcome the dimension curse of the conventional RL, deep reinforcement learning (DRL) is proposed, which uses deep learning and RL principles to create more efficient algorithms compared with RL, because deep neural networks are integrated with RL to fit the complex relationship between the action space and state space. Fig.~\ref{DRLframework} shows the structure of DRL for NOMA-MEC in remote-e-Health. In~\cite{Zhai2020iwc}, a DRL approach is proposed for MEC to deploy the computation services to the network edge, in which the users' request models and resource limitations is considered.

\subsection{Case study: NOMA-MEC in remote-e-Health}

We continue by providing case studies of NOMA-MEC in remote-e-Health. The major advantage of NOMA-MEC is to obtain speedy task computing taking advantages of both NOMA strategy and MEC. In this section, RL based solutions to sufficiently utilize the computing capabilities in NOMA-MEC network are studied.

In the proposed NOMA-MEC framework in e-Health, a long-term optimization problem is formulated that indicates a joint optimization of transmission decisions and computation resource allocation. Instead of solving a sophisticated joint optimization problem, we propose a modified RL algorithm, in which, the intelligent agent is capable of learning the optimal resource allocation scheme according to the history information and searching for the optimal trajectory automatically.

The performance of the proposed NOMA-MEC framework is shown in Fig.~\ref{offloading}. We can see from Fig.~\ref{offloading} that the proposed NOMA-MEC framework achieves lower energy consumption than the conventional OMA-MEC scheme. Also can be noted that the NOMA/OMA gain is higher with the increase of the task input size, which means that the benefit of NOMA is larger when we transmit larger computation tasks. In Fig.~\ref{offloading}, the computation tasks' MEC computing outperforms conventional local computing, because there are abound computing capabilities in the network edge. Also, we note that the proposed RL algorithm outperforms conventional schemes.

\section{CONCLUSION AND FUTURE DIRECTIONS}

\subsection{Concluding remarks}

In this paper, we have investigated the application of NOMA-enabled 5GB networks in remote-e-Health systems. Both NOMA-empowered ARs networks and NOMA-empowered EI networks are discussed by comparing to the conventional communication-based e-Health networks. A pair of case studies are provided to verify the performance of the proposed architecture, while the research challenges and future directions are also presented in terms of practical implementation.

\subsection{Challenges of integrating 3C into remote-e-Health}

\subsubsection{Cooperative NOMA-5GB design by exploiting heterogeneous mobility}

With the explosive development of mobile ARs in remote-e-Health systems, devices are expected to communicate with each other in an environment with heterogeneous mobility profiles. Nevertheless, there exist many mobility-related problems, which may become obstacles to the future 5GB networks, especially for the scenarios that massive devices access the wireless networks simultaneously. In practice, it is likely that the high-mobility users' channel conditions are worse than the low-mobility users' channel conditions, which is beneficial for the implementation of NOMA technique. However, how to reap the benefits of this channel difference is challenging.

\subsubsection{Pareto-optimization for meeting multiple objectives}

In contrast to the conventional wireless networks, NOMA-enabled 5GB in remote-e-Health systems are characterized by more rapidly fluctuating network topologies and more vulnerable communication links. Furthermore, NOMA-enabled 5GB rely on the seamless integration of heterogeneous network segments with the goal of providing improved QoS. Hence, it operates in a complex time-variant hybrid environment, where the classic mathematical models have limited accuracy~\cite{Wang2017taking}. Additionally, the challenging optimization problems encountered in remote-e-Health systems have to meet multiple objectives (delay, throughput, BER, power) in order to arrive at an attractive solution. However, determining the entire Pareto-front of optimal solutions is still challenging.

\subsection{Future directions}

\subsubsection{COVID-19 related applications}

COVID-19 pandemic has spread globally and resulted in an ongoing pandemic that not only has placed people at risk of illness and death, but also impacted the global economy and daily life. Quick individual screening is an effective method to protect citizens and make the best use of resources as the COVID-19 test kits are far from enough to provide nucleic acid testing for everyone and people may get infected anytime. Since fever is one of the common symptoms of being infected by COVID-19, identifying individuals with elevated body temperature (EBT) in public areas has been recognized as one of the most efficient methods to distinguish suspected infected individuals from the general public. Since the COVID-19 is highly infectious, it is expected to accomplish EBT detection missions without participation of trained staff or medical personnel.

By deploying thermal imaging empowered ARs in a particular public area for identifying individuals with high EBT, individuals who need additional screening can be detected and alerted. Thus, the spread of COVID-19 virus and infections in the public area can be slowed down or cut off dramatically. In addition, by integrating ARs instead of face-to-face detection, the staff or medical personnel can obtain people's body temperature data, as well as coordinate individuals with high EBT without approaching them in person for safety purposes, which could reduce the infection risk and save life of a large number of personnel working in public areas.

However, it is of the upmost importance to guarantee that all thermal imaging empowered ARs can move safely, which requires ultra-reliable and secure communication links between UAVs and their ground control stations (GCSs). Furthermore, high-rate and massive-connected NOMA ARs-BS communication links are also needed for sending back real-time thermal images from ARs to GCSs. To this end, a 5G-driven NOMA-ARs platform, which integrates ARs into the existing and future-generation wireless networks as new users, is a promising approach to tackle the aforementioned fundamental challenges.

\subsubsection{Federated learning for individuals' privacy preserving}

Since healthcare is a sensitive issue, medical data privacy preserving should be considered carefully. Federated learning (FL) is capable of achieving distributed network training utilizing large amount of local data, but update only network parameters instead of the sensitive raw data. The advantage of FL is that the central networks do not need local data of the patients for training, only the local networks' parameters are shared with the central networks. The local sensitive data of the patients is kept in the local storage units. Therefore, the privacy of the patients and hospitals are preserved while enabling excellent network training performance~\cite{Konecny2016nips}. According to the functions of FL, medical image classification (MIC) and medical big data prediction (MBDP) are also important application of EI in e-Health.

\begin{enumerate}

  \item MIC: MIC plays an important role in the computer version area. The goal of MIC is to recognize hidden information from medical image and provide fundamental information for doctors and researchers in disease diagnosis. Recent advancements in medical image collection and processing enable large volumes of medical images and medical data possible for complex network training. With the assist of FL, the neural network for MIC can be trained in a distributed manner. NOMA strategy can be adopted for parameters updating in FL based MIC, which provides a low-latency multiple access medical service to patients~\cite{Steven2006icml}.

  \item MBDP: The massive medical data collected from hospitals and health-care organization provide a new era for the medical services. The MBDP hold excellently values for the patients and hospitals. However, the complexity of clinic data require novel data processing techniques to extract useful features and information. High efficient AI technologies are needed for processing large volume of medical big data. The role of MBDP is to provide better health care services for individuals and communities.~\cite{chen2018icm}.

\end{enumerate}

 \end{document}